\documentclass[10pt,final,journal,letterpaper,twoside,twocolumn]{IEEEtran}
\usepackage{graphicx,cite,epsfig,amssymb,amsmath,subfigure,float}
\usepackage{graphicx,cite,epsfig}
\usepackage{amsmath,amssymb}
\usepackage{amsmath,bm}
\usepackage[noend]{algpseudocode}
\usepackage{algorithmicx,algorithm}
\usepackage{subfigure}
\usepackage{CJK}
\usepackage{color}

\begin{document}

\title{Ultra-Reliable Low-Latency Communications in Autonomous Vehicular Networks}

\author{\normalsize
Xiaohu Ge, ~\IEEEmembership{Senior~Member,~IEEE}\\

\thanks{\small{Xiaohu Ge is with the School of Electronic Information
and Communications, Huazhong University of Science and Technology, China
(Email: xhge@mail.hust.edu.cn).}}
}

\IEEEpubidadjcol
\maketitle


\date{\today}
\begin{abstract}
Autonomous vehicles are expected to emerge as a main trend in vehicle development over the next decade. To support autonomous vehicles, ultra-reliable low-latency communications (URLLC) is required between autonomous vehicles and infrastructure networks, e.g., fifth generation (5G) cellular networks. Hence, reliability and latency must be jointly investigated in 5G autonomous vehicular networks. Utilizing the Euclidean norm theory, we first propose a reliability and latency joint function to evaluate the joint impact of reliability and latency in 5G autonomous vehicular networks. The interactions between reliability and latency are illustrated via Monto-Carlo (MC) simulations of 5G autonomous vehicular networks. To improve both the reliability and latency performance and implement URLLC, a new network slicing solution that extends from resource slicing to service and function slicing is presented for 5G autonomous vehicular networks. The simulation results indicate that the proposed network slicing solution can improve both the reliability and latency performance and ensure URLLC in 5G autonomous vehicular networks.

\end{abstract}

\section{Introduction}
\label{sec1}
Vehicular networks are emerging as a key application scenario for fifth generation (5G) mobile communication systems \cite{1Alliance15}, and in the next decade, autonomous vehicles will represent one of the main transmitters/receivers of 5G vehicular networks \cite{Yang08}. Compared with traditionally manned vehicles, autonomous vehicles are extremely dependent on ultra-reliable low-latency communication (URLLC) in 5G vehicular networks. Previous studies have indicated that conflicts can occur between reliability and latency performance in vehicular networks \cite{2Ge17,3Liu16}. Hence, many studies have focused on optimizing either the reliability or latency performance in vehicular networks \cite{4Zaki11,5Ge16,6Cheng15}. Considering the URLLC requirement, a joint model of reliability and latency must be investigated for 5G autonomous vehicular networks. Moreover, developing a solution for improving both reliability and latency in 5G autonomous vehicular networks, i.e., implementing URLLC is a considerable challenge.

Studies have investigated methods of improving the latency or reliability of vehicular networks \cite{6Li17,7Liu17,Kato18,8Safiulin17,9Li15,10Sou11,11Shao15,17Yu16,20Sun16,16Wang16}. A mobile-edge computing (MEC) architecture was proposed for cellular vehicular networks in which an inter-cell handover mechanism was developed for vehicles to enhance the latency performance of mobility management \cite{6Li17}. Based on software-defined networks (SDNs) and OpenFlow technologies, a SDN-enabled network architecture assisted by MEC was proposed to reduce the latency in vehicular networks \cite{7Liu17,Kato18}. A multicast transmit beamforming technology was developed for vehicle to everything (V2X) communications that employs long-term evolution (LTE) multimedia broadcast single-frequency network (MBSFN) capabilities \cite{8Safiulin17}. To reduce the latency in multi-hop vehicular networks, a new scheme was proposed to optimize the one hop transmission range based on a genetic algorithm \cite{9Li15}. Considering a limited number of roadside units (RSUs) along roads, a routing scheme was proposed for broadcast-based safety applications with ultra-reliability in vehicular ad hoc networks (VANETs) \cite{10Sou11}. The connectivity probability was analyzed for platoon-based VANETs in which vehicles in the network have a Poisson distribution considering different traffic densities \cite{11Shao15}. The simulation results showed that the connectivity probabilities in VANETs based on platoons are larger than those in VANETs without platoons in the V2X communication scenarios. To improve the network capacity and system computing capability, a matrix game approach was developed to manage the cloudlet resources of vehicular networks \cite{17Yu16}. A three-stage radio resource management algorithm was developed to optimize resource sharing among vehicular users and cellular users in V2X applications \cite{20Sun16}. By formulating the virtual resource allocation and caching strategies as a joint optimization problem, a new framework with information-centric wireless virtualization and device-to-device communications was proposed to enable content caching not only in the air but also in mobile devices \cite{16Wang16}. However, enhancing both the reliability and latency of autonomous vehicular networks remains a key issue. To resolve this key issue, the network slicing technology is emerging as a potential solution for URLLC in autonomous vehicular networks.

Numerous network slicing concepts and methods have been proposed in the literatures on wireless networks \cite{6Rost17,11Foukas17,7Richart16,8Xu14,9Rost16,10Liu16,12Riggio16,13Samdanis16,14Liang15,15Liang15,21Zhang17}. The basic principles of network slicing that underly the mapping of dedicated and shared slices were discussed in \cite{6Rost17} considering the necessary flexibility and scalability associated with 5G network implementation. A common framework was presented to integrate and discussing the latest developments in 5G network slicing, and the identified gaps were evaluated \cite{11Foukas17}. Network slicing was introduced as an integral approach to wireless network virtualization and shown to promote the programmable and configurable characteristics of network services \cite{7Richart16}. A mobile-oriented Open-Flow protocol (MOFP) was proposed for the sharing of resources at adjacent base stations (BSs) and the implementation of service slicing for 5G cellular networks \cite{8Xu14}. In \cite{9Rost16}, network slicing technology was proposed to satisfy different network slicing requirements and content by dynamically scheduling the edges and central clouds in the studied wireless network. Based on network slicing technology, the virtual resource allocation of full-duplex relaying (FDR) networks was formulated as an optimization problem, and an efficient alternating direction method of multipliers (ADMM)-based distributed virtual resource allocation algorithm was developed to solve this problem \cite{10Liu16}. The virtual network functions (VNFs) placement problem was formalized for radio access networks, and a slice scheduling mechanism was proposed to ensure resource and performance isolation among different slices \cite{12Riggio16}. The concept of a 5G network slice broker was introduced for 5G networks, and it enables mobile virtual network operators, over-the-top providers and industry vertical market players dynamically to request and lease resources from infrastructure providers via signaling \cite{13Samdanis16}. In \cite{14Liang15}, the bandwidth of wireless networks was formed as a type of virtual slice by wireless network virtualization functions. Furthermore, \cite{15Liang15} presented a framework that enables wireless virtualization and discussed a number of challenges that must be addressed for the deployment of wireless virtualization in the next generation of mobile cellular networks. Network slicing technology has been introduced into vehicular networks by utilizing resource sharing schemes. Based on SDN and fog computing technologies, a new vehicular network architecture was proposed to improve the coverage probability of the network slicing technology in 5G vehicular networks \cite{21Zhang17}. However, investigations of URLLC in autonomous vehicular networks are surprisingly rare in the open literatures. Additionally, the impact of network slicing technology on the reliability and latency performance of vehicular networks has been limited to simple scenarios, and methods of improving both the reliability and latency based on network slicing have not been investigated in 5G autonomous vehicular networks.

Motivated by the above research gaps, a reliability and latency joint function is proposed in this paper to evaluate the joint impact of reliability and latency on 5G autonomous vehicular networks. Moreover, a new network slicing solution with service, function and resource slicing is proposed to improve both the reliability and latency of 5G autonomous vehicular networks. The contributions and novelties of this paper are summarized as follows.

\begin{enumerate}
\item Based on the Euclidean norm theory, a reliability and latency joint function is proposed to evaluate the joint impact of the reliability and latency on 5G autonomous vehicular networks. Moreover, the interactions between reliability and latency are analyzed for 5G autonomous vehicular networks, and a maximum of reliability and latency joint function with respect to the vehicle density is validated via Monto-Carlo (MC) simulations.
\item To improve both the reliability and latency performance of 5G autonomous vehicular networks, i.e., to implement URLLC, a new network slicing solution that extends from resource slicing to service and function slicing is presented. Moreover, a new network slicing algorithm is developed to improve both the reliability and latency performance of 5G autonomous vehicular networks.
\item The simulation results indicate that the proposed network slicing solution can improve both of the reliability and latency performance of 5G autonomous vehicular networks, i.e., by implementing URLLC. Moreover, the optimization of the proposed network slicing solution depends on the vehicle density in the 5G autonomous vehicular networks.
\end{enumerate}

The remainder of this paper is outlined as follows. Section II describes the system model of autonomous vehicular networks. Section III investigates the coupled relationship between the reliability and latency in 5G autonomous vehicular networks based on the proposed reliability and latency joint function. Additionally, the reliability and latency performance are analyzed via MC simulations for 5G autonomous vehicular networks. Section IV presents the proposed network slicing solution for improving the reliability and latency in 5G autonomous vehicular networks, in which network slicing is extended from network resource slicing to service and function slicing. Section V compares and analyzes the reliability and latency performance of 5G autonomous vehicular networks with and without network slicing. Finally, Section VI presents the conclusions.

\section{System Model}
\label{sec2}

\begin{figure*}
\vspace{0.1in}
\centering
\includegraphics[width=16cm,draft=false]{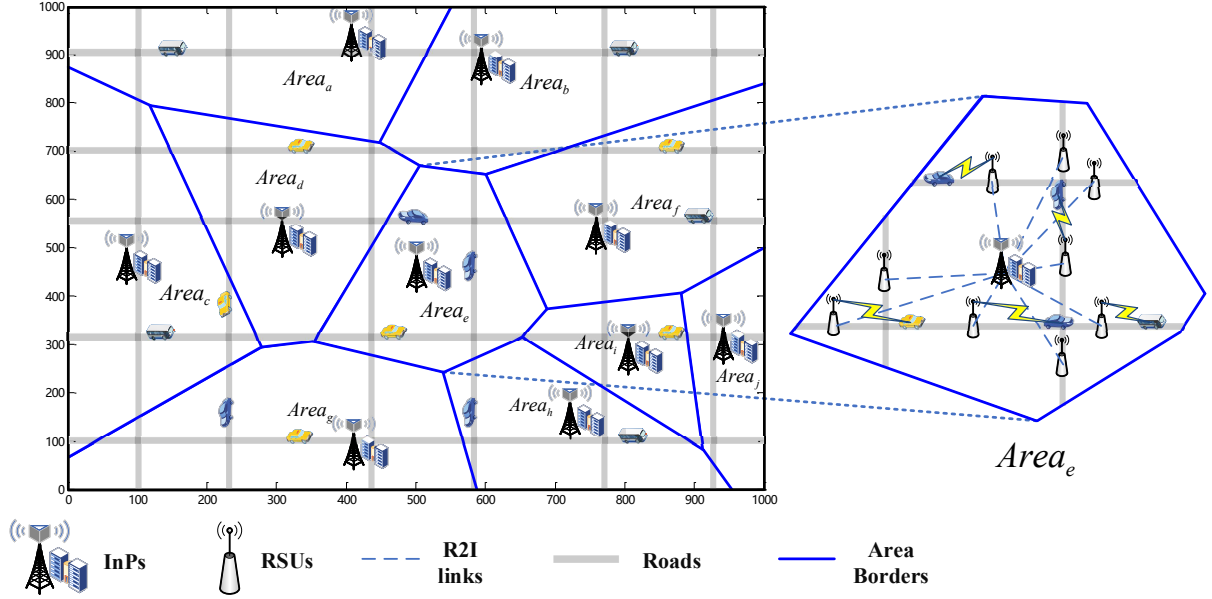}
\caption{\small System model. Coverage areas of different hot spots are shown as circles with different radius, ${s_{ij}}$  is the total number of AEs in circles marked as the purple regions.}
\end{figure*}
Without loss of generality, an urban outdoor environment is projected in a plane ${{\mathbb{R}}^{2}}$, in which the roads are modeled by the Manhattan Poisson line process (MPLP)\cite{24Valkeila08,25Klein97}. In Fig. 1, the road distribution model features two unit-density homogeneous Poisson point processes (PPPs) ${{\Psi }_{\text{x}}}$, ${{\Psi }_{\text{y}}}\subset {{\mathbb{R}}^{2}}$ along the $x\text{-axis}$ and $y\text{-axis}$, respectively. At each point in the processes, an avenue (west-east direction) and a street (north-south direction) grows infinitely along the $x\text{-axis}$ and $y\text{-axis}$, respectively. The resulting line process is denoted by $\mho $, which divides the plane into an infinite number of blocks. Based on the MPLP road model, the RSUs are assumed to be uniformly distributed along the roads in Fig. 1.

The infrastructure providers (InPs) $\mathcal{I}=\left\{ In{{p}_{a}},In{{p}_{b}},...,In{{p}_{i}} \right\}$ in the plane ${{\mathbb{R}}^{2}}$ are assumed to be governed by an independent PPP distribution. The plane ${{\mathbb{R}}^{2}}$ is split into irregular polygons that correspond to different cell coverage areas. The split method is based on the Delaunay Triangulation method where the perpendicular bisector lines are connected by each pair of InPs \cite{26Ge15}. This stochastic and irregular topology forms a so-called Poisson-Voronoi tessellation (PVT) \cite{27Mattfeldt96}. In each cell coverage area, an InP is located in the center of the cell and connects with all RSUs inside this cell coverage area. The InP gathers information from the RSUs and provides a virtualized network platform (VNP) to share network resources with the RSUs in the PVT cell.

Based on the Palm theory \cite{27Mattfeldt96,28zhong17}, the analytical results for a typical PVT cell, e.g., $\text{Are}{{\text{a}}_{e}}$ in Fig. 1, can be extended to the entire PVT network. Hence,  $\text{Are}{{\text{a}}_{e}}$ is magnified in Fig. 1 to show the distribution details in a typical PVT cell. In $\text{Are}{{\text{a}}_{e}}$, the RSUs have network resources that can be used to communicate with autonomous vehicles on the road by millimeter wave wireless transmissions. Moreover, the beamforming technology is adopted for wireless links between the RSUs and vehicles to avoid the interference from adjacent RSUs. For each RSU, a RSU to infrastructure provider (R2I) link is configured to provide communications between the RSU and InP.

\section{Reliability and Latency Joint Function in 5G Autonomous Vehicular Networks}
Most studies have separately modeled and analyzed the reliability and latency performances of vehicular networks \cite{Yang09}; however, these parameters are coupled in vehicular networks. When an algorithm is applied to optimize the latency or reliability performance of vehicular networks, the other performance parameters in the vehicular network will be affected, and performance deterioration can occur \cite{Yang2007}. Considering the URLLC requirement in 5G autonomous vehicular networks, the interactions between reliability and latency must be modeled and analyzed. In this section, a new reliability and latency joint function based on the Euclidean norm theory is proposed for 5G autonomous vehicular networks. Furthermore, the coupled relationship between reliability and latency is analyzed for 5G autonomous vehicular networks.

\subsection{Reliability and Latency Joint Function}

Based on the Euclidean norm theory, the reliability and latency joint function of 5G autonomous vehicular networks $\mathbb{S}\left( {{L_t},{L_p}} \right)$ is defined as follows:

\[\begin{gathered}
  \mathbb{S}\left( {{L_t},{L_p}} \right) = {\left( {{{\left| {{L_t}} \right|}^2} + {{\left| {\omega  \cdot {L_p}} \right|}^2}} \right)^{{1 \mathord{\left/
 {\vphantom {1 2}} \right.
 \kern-\nulldelimiterspace} 2}}} \hfill \\
   = {\left( {{{\left| {{I_t}\left( {{T_{req}}} \right)} \right|}^2} + {{\left| {\omega  \cdot {I_p}\left( {{P_{req}}} \right)} \right|}^2}} \right)^{{1 \mathord{\left/
 {\vphantom {1 2}} \right.
 \kern-\nulldelimiterspace} 2}}} \hfill \\
\end{gathered} \, \tag{1}\]
where ${L_p}$ and ${L_t}$ are the reliability and latency utilities in autonomous vehicular networks, respectively; $\omega $ is the weight factor that balances the effects of reliability and latency on $\mathbb{S}\left( {{L_t},{L_p}} \right)$; ${I_p}\left( {{P_{req}}} \right)$ is the reliability utility function with the reliability constraint ${P_{req}}$; and ${I_t}\left( {{T_{req}}} \right)$  the latency utility function with the latency constraint ${T_{req}}$, $\left| \cdot \right|$ is the norm operation. Based on the reliability and latency joint function, both the reliability and latency performance can be evaluated by the same metric, i.e., the norm length in Euclidean space. Therefore, the URLLC can be investigated based on a uniform model of the reliability and latency joint function in 5G autonomous vehicular networks.

To evaluate how reliability and latency are satisfied in autonomous vehicular networks, the reliability and latency utility functions are extended as follows:
\[{I_p}\left( {{P_{req}}} \right) = {e^{{a_P}\left( {\frac{{P - {P_{req}}}}{{{P_{req}}}}} \right)}},\tag{2}\]

\[{I_t}\left( {{T_{req}}} \right) = {e^{{a_T}\left( {\frac{{{T_{req}} - T}}{{{T_{req}}}}} \right)}},\tag{3}\]
where $P$ and $T$ are the reliability and latency values in 5G autonomous vehicular networks, respectively; ${P_{req}}$ and ${T_{req}}$ are the required reliability and latency thresholds for the desired services in 5G autonomous vehicular networks, respectively; and ${{a}_{P}}$ and ${{a}_{T}}$ are the weighted factors for the reliability and latency utility functions, respectively. When the values of reliability and latency are less than the required thresholds in autonomous vehicular networks, the reliability and latency utility function values are always larger than 0 and less than 1, i.e., $0<{{I}_{t}}\left( {{T}_{req}} \right)<1$ and $0<{{I}_{p}}\left( {{P}_{req}} \right)<1$. When the reliability and latency requirements are satisfied in autonomous vehicular networks, i.e., the reliability and latency values are larger than or equal to the required thresholds, the reliability and latency utility function values are always larger than or equal to 1, i.e., ${{I}_{t}}\left( {{T}_{req}} \right)\ge 1$ and ${{I}_{p}}\left( {{P}_{req}} \right)\ge 1$. Moreover, the reliability utility function value increases as the reliability increases, and the latency utility function value increases as the latency in autonomous vehicular networks decreases. In this case, the value of ${\mathbb{S}\left( {{L_t},{L_p}} \right)}$ increases as the reliability and latency improve in 5G autonomous vehicular networks. When the value of ${\mathbb{S}\left( {{L_t},{L_p}} \right)}$ is less than 1, either the reliability or latency requirement has not been satisfied in the 5G autonomous vehicular networks.

\subsection{Reliability Model}

The unreliability of messages in 5G autonomous vehicular networks is mainly caused by transmission errors in wireless links between the vehicles and RSUs. In this case, the reliability of autonomous vehicular networks is denoted by the probability of the message being successfully transmitted between the vehicle and the RSU. The wireless link between the vehicle and RSU is denoted by $\mathbb{L}$, and the number of hops is ${{\text{N}}_{\mathbb{L}}}$. The distance between two nodes in the wireless link $\mathbb{L}$ is denoted by $D_{i,j}^{x}$, $x \in \left\{ {V,R} \right\}$, where $i$ and $j$ are positive integers, $D_{i,j}^{V}$ denotes the distance between two vehicles and $D_{i,j}^{R}$ denotes the distance between the vehicle and the RSU. If the vehicle node is denoted as $D_i^{V}$ and the RSU node is denoted as $D_j^{R}$, then the wireless link $\mathbb{L}$ between the vehicle $D_i^{V}$ and RSU $D_j^{R}$ is expressed as a set as follows:
\[\mathbb{L}\text{=}\left\{ D_{i,1}^{V},D_{1,2}^{V},\cdots ,D_{n,j}^{R} \right\},\tag{4}\]
where $n$ is a positive integer.

In this paper, millimeter wave transmissions are adopted for the wireless links in 5G autonomous vehicular networks, and the frequency of the millimeter wave transmissions is 75 GHz \cite{33Ge18,25Ghosh14}. Without loss of generality, every RSU is assumed to cover a road with a length of $L$ meters in an urban environment. The density of RSUs in an urban area is denoted as ${{\rho }_{RSU}}=\frac{1}{L}$. The vehicle density on the road is configured as ${{\rho }_{i}}$. When a vehicle ${{V}_{a}}$ is covered by $RS{{U}_{i}}$, the distance $d$ between the vehicle and the RSU is assumed to be governed by a uniform distribution, i.e., $d\in U\left( 0,L \right)$, with the expectation that $E\left( d \right)={}^{L}/{}_{2}$. Based on the measured results in \cite{25Ghosh14,Kato14}, the path loss $PL[dB](\cdot)$ between the vehicle ${{V}_{a}}$ and the RSU $RS{{U}_{i}}$ can be expressed as follows:
\[PL[dB](d) = 69.6 + 20.9\log (d) + \xi ,\xi  \sim \left( {0,{\sigma ^2}} \right),\tag{5}\]
where $\xi$ is additive white Gaussian noise (AWGN) with a mean of 0 and variance $\sigma ^2$.

In this paper, beamforming technology is adopted for millimeter wave transmissions. Considering the directionality and fast fading features of millimeter wave transmissions, the interference among vehicles and RSUs is ignored in 5G autonomous vehicular networks. The threshold of the signal-to-noise ratio (SNR) at the receivers is configured as $\theta$ in one hop in a 5G autonomous vehicular network. When the SNR at the receivers is larger than or equal to the threshold $\theta$, the signal can be successfully accepted at the receivers \cite{Xiaohu18,Xiaohu19}. The one hop successful transmission probability $P_{hop}$ that the message can be successfully transmitted in a time slot is as follows \cite{Bin18}:
\[{P_{hop}} = {\rm P}\left( {L \le {P_{tx}}(dB) - \theta (dB) - {N_0}{W_{mmWave}}(dB)} \right),\tag{6}\]
where ${{P}_{tx}}$ is the transmission power, ${{N}_{0}}$ is the power spectrum density of AWGN, and ${{W}_{mmWave}}$ is the bandwidth of the millimeter wave transmissions. By substituting (5) into (6), and the one hop successful transmission probability $P_{hop}$ can be expressed as follows:
\[\begin{array}{*{20}{l}}
  \begin{gathered}
  {P_{hop}} = {\text{P}}\left( {\xi  \leqslant {P_{tx}}(dB) - \theta (dB)} \right. \hfill \\
  \left. {{\kern 1pt} {\kern 1pt} {\kern 1pt} {\kern 1pt} {\kern 1pt} {\kern 1pt} {\kern 1pt} {\kern 1pt} {\kern 1pt} {\kern 1pt} {\kern 1pt} {\kern 1pt} {\kern 1pt} {\kern 1pt} {\kern 1pt} {\kern 1pt} {\kern 1pt} {\kern 1pt} {\kern 1pt} {\kern 1pt} {\kern 1pt} {\kern 1pt} {\kern 1pt} {\kern 1pt} {\kern 1pt} {\kern 1pt} {\kern 1pt} {\kern 1pt} {\kern 1pt} {\kern 1pt} {\kern 1pt}  - {N_0}{W_{mmWave}}(dB) - 69.6 - 20.9{{\log }_{10}}d} \right) \hfill \\
\end{gathered}  \\
  {\quad \quad  = \frac{1}{2}\left( {1 + erf\left( {\frac{{\psi \left( d \right)}}{{\sqrt 2 \sigma }}} \right)} \right)}
\end{array},\tag{7}\]
where $erf(\centerdot )$ is the deviation function and $\psi \left( d \right)={{P}_{tx}}(dB)-\theta (dB)-{{N}_{0}}{{W}_{mmWave}}(dB)-69.6-20.9{{\log }_{10}}d$.

Considering the multi-hop process in a wireless link, the reliability of 5G autonomous vehicular networks is calculated as follows:
\[{\rm P} = \prod\limits_{D_{i,j}^x \in \mathbb{L}} {{P_{hop}}\left( {D_{i,j}^x} \right)} .\tag{8}\]

\subsection{Latency Model}

The messages between vehicles and adjacent RSUs are transmitted by vehicle to infrastructure (V2I) links. The RSU normalizes all received messages and forms a message queue to manage all vehicle messages in the coverage area. Without loss of generality, the total latency between the vehicles and RSUs can be divided into propagation latency and handling latency. Hence, the total latency of message $T$ can be expressed as follows:

\[T=Tt_{_{M}}^{p}+Tt_{_{M}}^{q},\tag{9}\]
where $Tt_{_{M}}^{p}$ is the propagation latency in the wireless links and $Tt_{_{M}}^{q}$ is the handling latency in the queues of RSUs.

When the wireless link between the vehicle and the RSU comprises one wireless hop, the propagation latency is expressed as follows:
\[{{T}_{hop}}\left( d \right)={}^{{{t}_{slot}}}/{}_{{{P}_{hop}}}=\frac{2{{t}_{slot}}}{1+erf\left( \frac{\psi \left( d \right)}{\sqrt{2}\sigma } \right)},\tag{10}\]
where ${{t}_{slot}}$ is a constant time slot used for transmitting messages.

When the wireless link $\mathbb{L}$ between the vehicle and the RSU is composed of multiple hops, i.e., a message from the vehicle is relayed via multiple vehicles to reach the desired RSU, the propagation latency of multi-hop $Tt_{_{M}}^{p}$ in the wireless link $\mathbb{L}$ is expressed as follows:
\[\begin{array}{l}
Tt_{_M}^p = {T_{trans}}{\rm{ + }}{T_{proc}}\\
\quad\quad = \sum\limits_{D_{i,j}^x \in \mathbb{L}} {{T_{hop}}\left( {D_{i,j}^x} \right)}  + \left( {{{\rm{N}}_\mathbb{L}} - 1} \right){t_{proc}}
\end{array},\tag{11}\]
where ${{T}_{trans}}$ is the transmission latency in the wireless link $\mathbb{L}$ and ${{T}_{proc}}$ is the total relaying latency, which includes the processing latency ${{t}_{proc}}$ of every relaying vehicle in the wireless link $\mathbb{L}$.

Every RSU is configured with the same number of local resource blocks $R{{B}_{rsu}}$. Considering the random mobility of vehicles on the roads, the number of vehicles in the coverage area of every RSU is different, as is the arrival rate of messages at every RSU. In this case, the message handling process at the RSU is assumed as a $GI/M/1/\infty $ queuing system. In this $GI/M/1/\infty $ queuing system, it is assumed that ${{\tau }_{0}}=0$ is the arrival epoch of the first task; the inter-arrival times $\left\{ {{\tau }_{i}},i\ge 1 \right\}$ are independent and identically distributed with a general distribution function denoted by $F\left( t \right),t\ge 0$; and the mean is $0<\frac{1}{\lambda }=\int_{0}^{\infty }{tdF\left( t \right)},\left( \lambda >0 \right)$, where $\lambda $ is the arrival rate. The service times $\left\{ {{\chi }_{i}},i\ge 1 \right\}$ during a service period are exponentially distributed at the rate of $\mu \left( \mu >0 \right)$; therefore, the distribution function of service times is denoted by $G\left( t \right)=1-{{e}^{-\mu t}},t\ge 0$. Considering the $GI/M/1/\infty $ queuing system, the average service time of a message at RSUs is dependent on the number of resource blocks in the RSU \cite{40Baba05}. To simplify the analysis, $\frac{1}{{{\mu }_{0}}}$ denotes the average service time when a message is serviced by a resource block, and $\frac{1}{{R{B_{rsu}}\left( i \right) \cdot {\mu _0}}}$ is the average service time when a message is serviced by $RS{{U}_{i}}$ with $R{{B}_{rsu}}\left( i \right)$ resource blocks. In the $GI/M/1/\infty $ queuing system, the handling latency of a message at an RSU corresponds to the following theorems.

\textbf{Theorem 1}: Without loss of generality, $RS{{U}_{i}}\left( i=1,2,...,m \right)$ covers a road with a length of $L$ meters. The average density of vehicles on the road is denoted as ${{\rho }_{i}}$, and the transmission rate of messages generated by a vehicle is ${{\lambda }_{s}}$. The cumulative distribution function (CDF) of the message interval arrival time at $RS{{U}_{i}}$ is expressed as follows:
\[F\left( t \right) = 1 - \frac{{{e^{ - {\rho _i}L\left( {1 - {e^{ - {\lambda _s}t}}} \right)}} - {e^{ - {\rho _i}L}}}}{{1 - {e^{ - {\rho _i}L}}}}.\tag{12}\]
The proof for \textbf{Theorem 1} is provided in Appendix A.

\textbf{Theorem 2}: When the message handling process at an RSU is assumed to occur based on a $GI/M/1/\infty $ queuing system, the distribution of the message dwelling time at the RSU is governed by the following equation:
\[W\left( t \right) = 1 - {e^{ - \mu t}} + \frac{{{K^*}}}{{1 - \delta }}\left[ {1 - {e^{ - c\mu \left( {1 - \delta } \right)t}}} \right],\tag{13}\]
with
\[{K^ * } = {\left[ {\frac{1}{{1 - \delta }} + \sum\limits_{k = 1}^c {\left( {\frac{{\left( {\begin{array}{*{20}{c}}
c\\
k
\end{array}} \right)}}{{{D_k}\left( {1 - {\varepsilon _k}} \right)}} \cdot \frac{{c\left( {1 - {\varepsilon _k}} \right) - k}}{{c\left( {1 - \delta } \right) - k}}} \right)} } \right]^{ - 1}},\tag{14}\]

\[{D_k} = \left\{ {\begin{array}{*{20}{c}}
1&{k = 0}\\
{\prod\limits_{l = 1}^k {\frac{{{\varepsilon _l}}}{{1 - {\varepsilon _l}}}} }&{k = 1,2, \cdots ,c}
\end{array}} \right.,\tag{15}\]

\[\begin{array}{*{20}{c}}
{{\varepsilon _l} = \int_0^\infty  {{e^{ - l\mu t}}} dF\left( t \right),}&l
\end{array} = 1,2, \cdots.\tag{16}\]
The proof for \textbf{Theorem 2} is provided in Appendix B.

Furthermore, the average dwelling time of a message in the $GI/M/1/\infty $ queuing system is derived as follows:
\[Tt_{_S}^q = E\left[ {W\left( t \right)} \right]{\rm{ = }}\frac{1}{\mu }{\rm{ + }}\frac{{{K^ * }}}{{c\mu {{\left( {1 - \delta } \right)}^2}}}.\tag{17}\]
Considering the $GI/M/1/\infty $ queuing system at the RSU, the parameter $c$ is set as a constant, i.e., $c=1$. Hence, the average dwelling time of a message is derived as follows:
\[Tt_{_S}^q = E\left[ {W\left( t \right)} \right]{\rm{ = }}\frac{1}{{\mu \left( {1 - \delta } \right)}}.\tag{18}\]
When the value of $\delta $ is limited as $\delta \in \left( 0,1 \right)$, $\delta $ can be solved based on the following relation:
\[\sum\limits_{k = 0}^\infty  {{\delta ^k} \cdot \int_0^\infty  {{e^{ - \mu t}}\frac{{{{\left( {\mu t} \right)}^k}}}{{k!}}} } dF\left( t \right) = \int_0^\infty  {{e^{ - \mu \left( {1 - \delta } \right)t}}dF\left( t \right)}  = \delta .\tag{19}\]
When the CDF of message interval time in (12) is substituted into (19), (19) can be extended as follows:

\[\int_0^\infty  {{e^{ - \mu \left( {1 - \delta } \right)t}}\frac{{{\lambda _s}{\rho _i}L{e^{ - {\rho _i}L + {\rho _i}L{e^{ - {\lambda _s}t}} - {{\lambda _s}t}}}}}{{1 - {e^{ - {\rho _i}L}}}}dt}  = \delta.\tag{20}\]

\subsection{Performance Analysis of Reliability and Latency}

To analyze the reliability and latency performance of 5G autonomous vehicular networks, the default simulation parameters are configured as follows: the weight factor $\omega $ is set to 10, the road length $L$ covered by the RSU is 100 meters \cite{32Sou10,33Sun16}, the vehicle density on the road ${{\rho }_{i}}$ is 0.2 vehicle per meter (veh/m) \cite{34Liu16}, the transmission rate of messages generated from a vehicle ${{\lambda }_{m}}$ is 50 messages per second, the average service time ${1}/{{{\mu }_{0}}}\;$ is 5 milliseconds \cite{35Anttonen14}, the transmission power of the vehicle ${{P}_{tx}}$ is 30 dBm \cite{36Xiang13}, the noise power density ${{N}_{0}}$ is -174 dBm/Hz \cite{36Xiang13}, the duration of a slot ${{t}_{slot}}$ is 50 microseconds \cite{37Li15}, the SNR threshold $\theta $ is 5 dB \cite{36Xiang13}, and the number of resource blocks in a RSU $R{{B}_{rsu}}$ is 10. Without loss of generality, the wireless link between the vehicle and RSU is composed of random multiple hops, which are governed by a uniform distribution. Moreover, MC simulations are used for the performance analysis in this paper.

\begin{figure}
\vspace{0.1in}
\centerline{\includegraphics[width=9.5cm,draft=false]{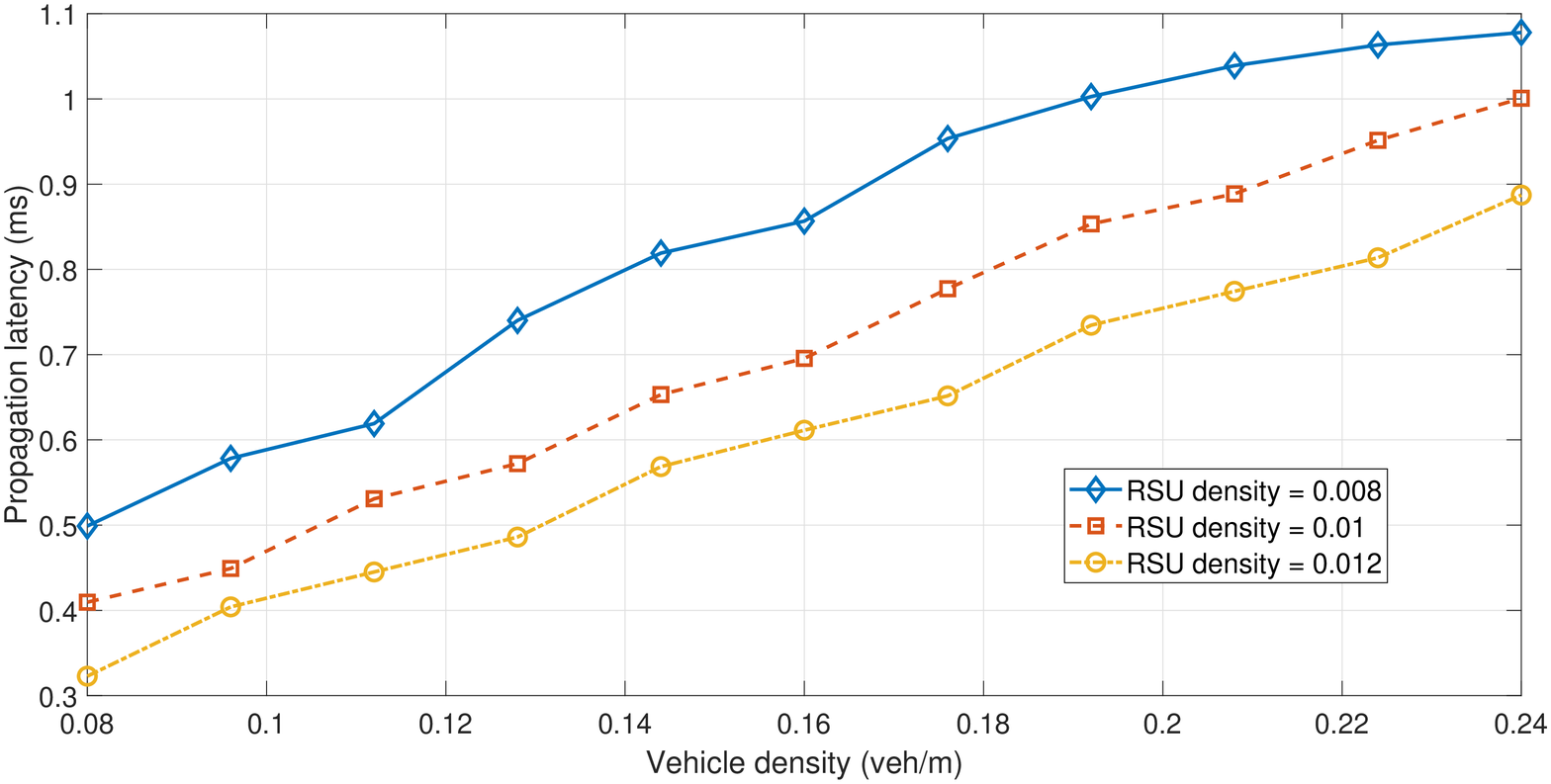}}
\caption{\small Propagation latency with respect to the vehicle density based on different RSU densities.}
\end{figure}

\begin{figure}
\vspace{0.1in}
\centerline{\includegraphics[width=9.5cm,draft=false]{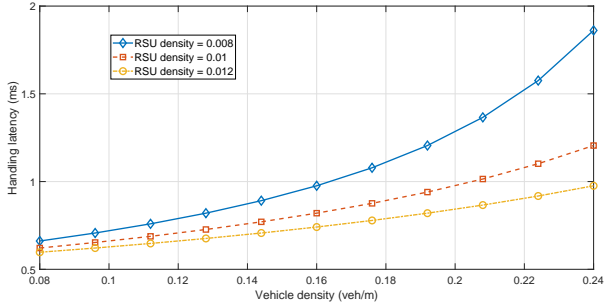}}
\caption{\small Handling latency with respect to the vehicle density considering different RSU densities.}
\end{figure}

Fig. 2 describes the propagation latency with respect to the density of vehicles considering different RSU densities. When the RSU density is fixed, the propagation latency increases with the vehicle density. When the vehicle density is fixed, the propagation latency decreases as the RSU density increases.

Fig. 3 shows the handling latency with respect to the density of vehicle considering different RSU densities. When the RSU density is fixed, the handling latency increases as the vehicle density increases. When the vehicle density is fixed, the handling latency decreases as the RSU density increases.

\begin{figure}
\vspace{0.1in}
\centerline{\includegraphics[width=9.5cm,draft=false]{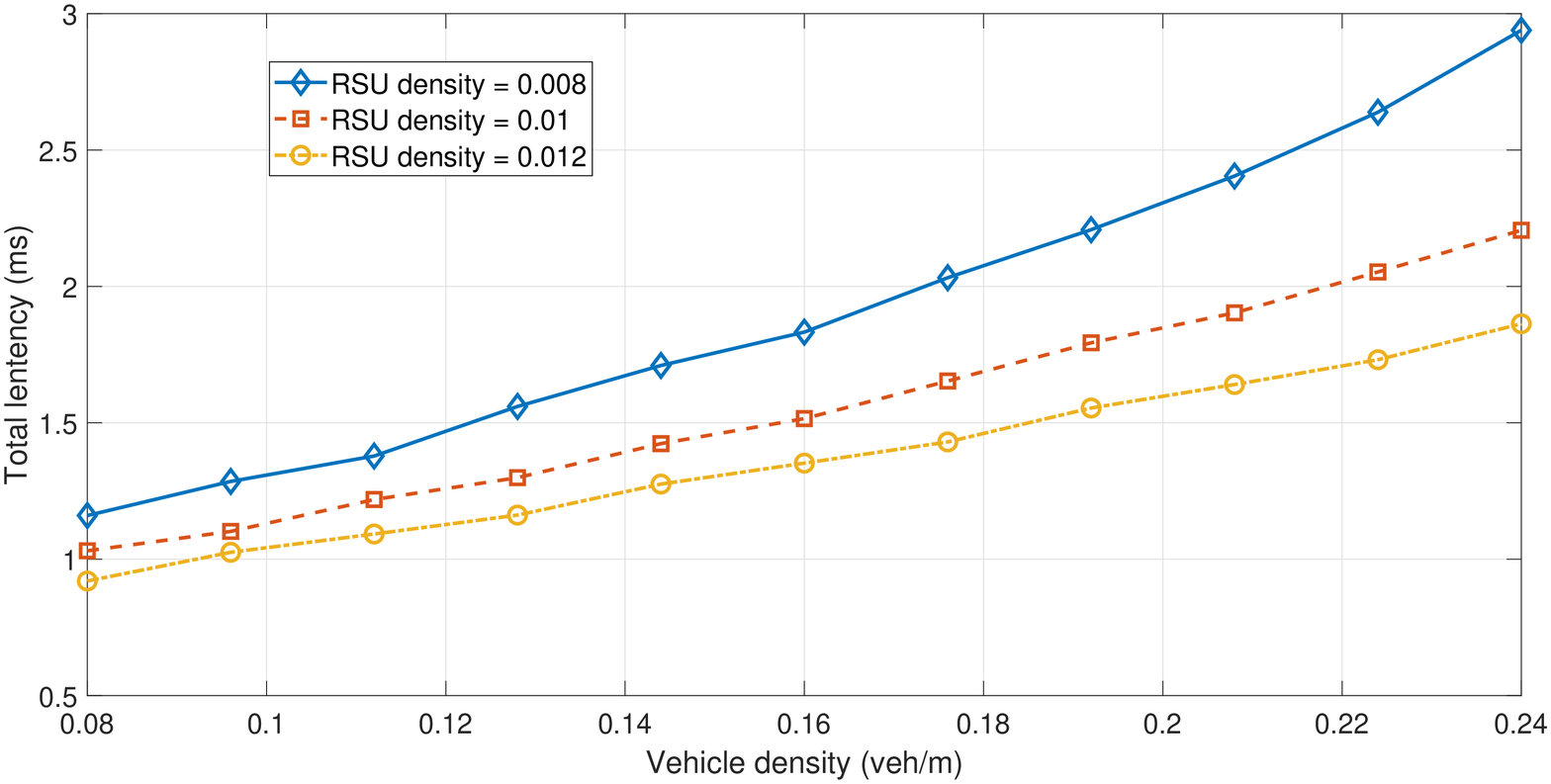}}
\caption{\small Total latency with respect to the vehicle density considering different RSU densities.}
\end{figure}

Fig. 4 illustrates the total latency with respect to the density of vehicle considering different RSU densities. When the RSU density is fixed, the total latency increases as the vehicle density increases. When the vehicle density is fixed, the total latency decreases as the RSU density increases.

\begin{figure}
\vspace{0.1in}
\centerline{\includegraphics[width=9.5cm,draft=false]{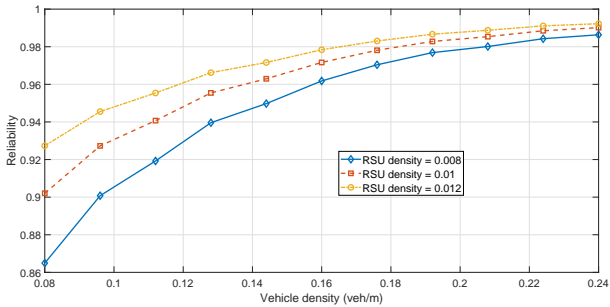}}
\caption{\small Reliability with respect to the vehicle density considering different RSU densities.}
\end{figure}

\begin{figure}
\vspace{0.1in}
\centerline{\includegraphics[width=9.5cm,draft=false]{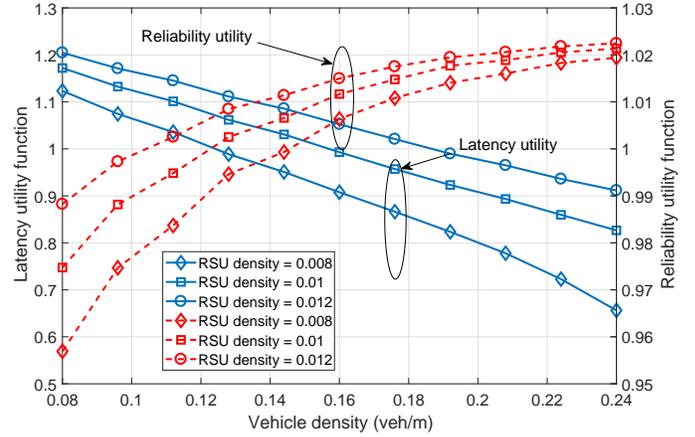}}
\caption{\small Relationship between the reliability and latency utility functions in 5G autonomous vehicular networks.}
\end{figure}

\begin{figure}
\vspace{0.1in}
\centerline{\includegraphics[width=9.5cm,draft=false]{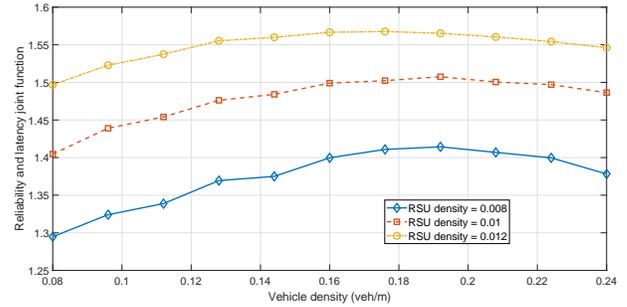}}
\caption{\small Reliability and latency joint function with respect to the vehicle density considering different RSU densities.}
\end{figure}

Fig. 5 depicts the reliability with respect to the density of vehicle considering different RSU densities. When the RSU density is fixed, the reliability increases as the vehicle density increases. When the vehicle density is fixed, the reliability increases as the RSU density increases.

Fig. 6 indicates the relationship between the reliability and latency utility functions in 5G autonomous vehicular networks. When the RSU density is fixed, the reliability utility function increases and the latency utility function decreases as the vehicle density increases. Hence, a conflict occurs between the reliability and latency performance with respect to the vehicle density in 5G autonomous vehicular networks. When the vehicle density is fixed, the reliability and latency utility functions increase as the RSU density increases.

Fig. 7 analyzes the reliability and latency joint function with respect to the vehicle density considering different RSU densities. When the RSU density is fixed and the vehicle density is less than 0.19 veh/m, the values of the reliability and latency joint function initially increase as the vehicle density increases. When the RSU density is fixed and the vehicle density is larger than or equal to 0.19 veh/m, the values of the reliability and latency joint function decrease as the vehicle density increases. Hence, a maximum of reliability and latency joint function value exists with respect to the vehicle density. When the vehicle density is fixed, the values of the reliability and latency joint function value increase as the RSU density increases.

\section{Network Slicing Solution}

Based on the results in Fig. 6, a conflict is observed regarding the reliability and latency performance in 5G autonomous vehicular networks. To implement URLLC, both the reliability and latency performance in 5G autonomous vehicular networks must be improved. Considering the reliability and latency gains associated with using network slicing technology, network slicing has emerged as an attractive solution for 5G autonomous vehicular networks.
In conventional network slicing technology, only network resources are sliced to improve the resource utilization in wireless networks. However, satisfying the URLLC requirement in 5G autonomous vehicular networks based on network resource slicing  alone is difficult. In this case, we extend network slicing from resource slicing to service and function slicing to improve both the reliability and latency of 5G autonomous vehicular networks. The relevant methods are described in detail in this section.

\subsection{Service Slicing}

Different types of vehicular services are available in 5G autonomous vehicular networks. To improve the service access efficiency in 5G autonomous vehicular networks, we normalize different types of vehicular services into three types of service slices.

\begin{itemize}
\item State-Report Service Slice

State-report service slices (SRSSs) are used to provide the state of the vehicle, such as the speed and location. SRSSs have the following features: the arrival rate is high; the holding time of the state is short; and the SRSSs have similar contexts. SRSSs are used for vehicle collision avoidance, vehicle lane changes and vehicle deceleration. Hence, SRSSs must be quickly accessed. Moreover, SRSSs should be processed at RSU locations to reduce the response time.

\item Event-Driven Service Slice

Event-driven service slices (EDSSs) are generated by specific events, such as the emergency broadcasts and the road information updates. Although the EDSS arrival rate is low, EDSSs usually involve with security issues. Hence, EDSSs must be accessed with a low latency and high reliability in 5G autonomous vehicular networks.

\item Entertainment-Application Service Slice

Entertainment-application service slices (EASSs) are generated from vehicle users. The latency requirement of EASSs depends on the type of service applications. Different types of service applications have different latency requirements. Hence, the EASS requirements change for different types of service applications in 5G autonomous vehicular networks.

\end{itemize}

Based on the analysis of the three types of service slices, different types of service slices can be distinguished by the arrival rate, handling time and latency requirement. To simplify the analysis of service slices in 5G autonomous vehicular networks, the service slice of a 5G autonomous vehicular network is denoted as  ${\rm S}\left\{ {{\lambda _a},{\mu _s},{T_{req}}} \right\}$, where ${\lambda _a}$ is the arrival rate of the service slice, ${\mu _s}$ is the handling time of service slice and ${T_{req}}$ is the latency requirement of the service slice.

\subsection{Function Slicing and Resource Slicing}

\begin{figure*}
\vspace{0.1in}
\centering
\includegraphics[width=16cm,draft=false]{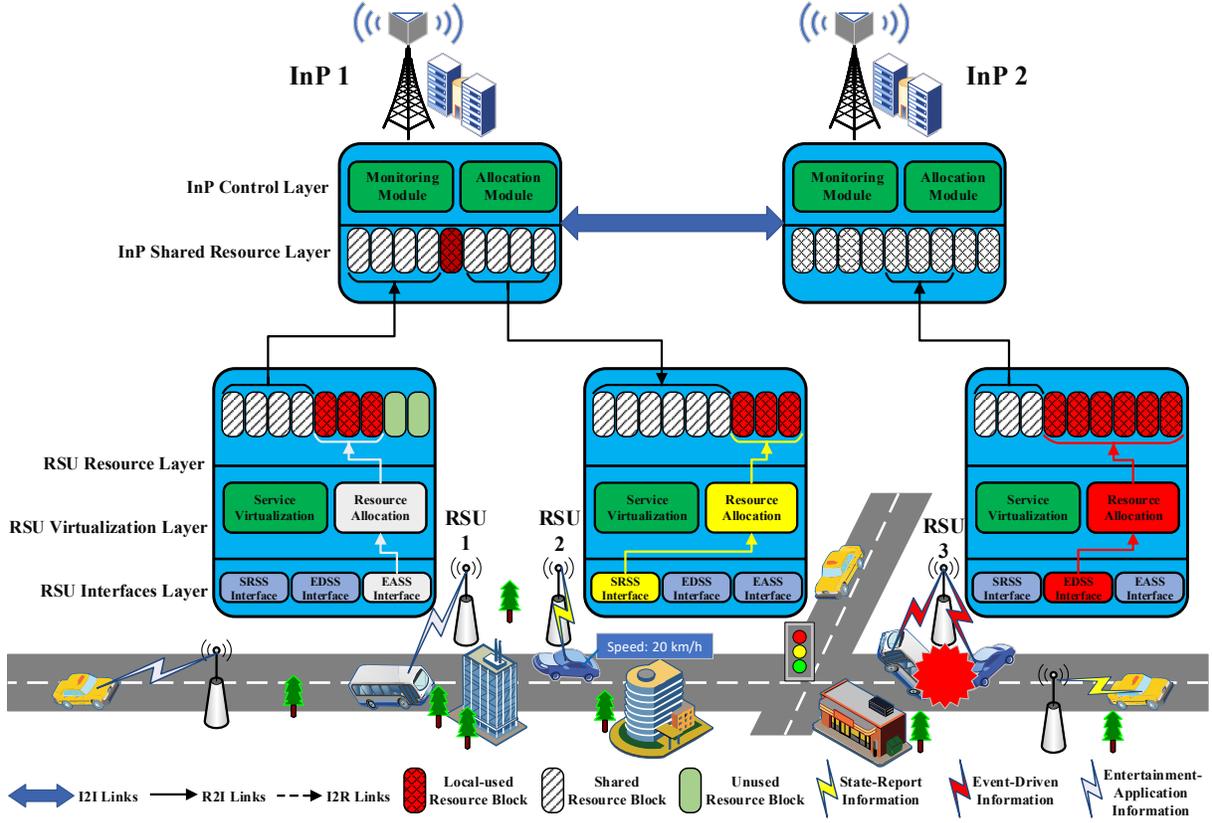}
\caption{\small Autonomous vehicular network architecture.}
\end{figure*}

To improve both the reliability and latency, the protocol functions at the RSUs and InPs must be sliced and re-structured by SDN technologies to support service and resource slicing in 5G autonomous vehicular networks.
Therefore, we propose a new 5G autonomous vehicular network architecture to implement service, resource and function slicing, as shown in Fig. 8. In this paper, function slicing is implemented by dynamically scheduling and assembling function modules in the logical layers of the proposed autonomous vehicular network architecture.
Based on the proposed autonomous vehicular network architecture, the protocol structure of an InP is composed of two logical layers, i.e., the InP control layer and InP shared resource layer. The InP control layer includes two function modules: the monitoring module and the allocation module. The monitoring module monitors the number of occupied resource blocks at the RSUs in an InP coverage area. Based on the information from the monitoring module, such as the number of occupied resource blocks that reach the available number of resource blocks and the number of resource blocks that is still required by the service slices at the RSU, the allocation module will schedule the unoccupied resource blocks from adjacent RSUs to support the requirements of the specified RSUs. The InP shared resource layer is a type of virtual network resource pool that includes all unoccupied resource blocks in the coverage of the InP. The resource slicing is performed by the InP shared resource layer to improve the resource utilization efficiency in 5G autonomous vehicular networks.
In the real world, resource blocks are stored at the RSUs. According to the logic relationships in the proposed autonomous vehicular network architecture, all unoccupied resource blocks in the coverage of the InP can be regarded as stored in the virtual network resource pool in the InP shared resource layer. Furthermore, all resource blocks stored in the virtual network resource pool can be sliced to effectively satisfy the resource requirements of service slices at the RSUs.

The protocol structure of an RSU is divided into three logical layers: the RSU interface layer, RSU virtualization layer and RSU resource layer. The RSU interface layer provides different interfaces corresponding to different types of service slices. The main function of the RSU interface layer is to separate RSU resource requirements with different types of service slices. Based on the RSU interface layer, the relationships among different types of service slices are normalized by separating the resource requirements.

The function of the RSU virtualization layer is to allocate locational RSU resource blocks to satisfy the requirements of the service slices from the RSU interface layer. Moreover, the RSU virtualization layer can optimize the handling process to promptly respond to the service slices with low latency constraint.

The RSU resource layer manages the resource blocks at the RSU location. The RSU resource blocks were  previously allocated for the location service slices. Thus, the RSU virtualization layer can require resource blocks from the associated InP control layer when the RSU resource layer cannot satisfy the requirements of the location service slices. When the resource blocks from the InP shared resource layer are scheduled in the RSU virtualization layer, the RSU resource layer will allocate the additional resource blocks from the RSU virtualization layer to meet the requirements of the location service slices.

\subsection{Network Slicing Algorithm}
Assuming that ${{N}_{RSU}}$ RSUs exist in a coverage area of an InP, the average number of ${{N}_{RSU}}$ is derived in Appendix C. A ${{N}_{RSU}}*{{N}_{RSU}}$ matrix ${{\bf{{\rm A}}}_{{N_{RSU}}*{N_{RSU}}}}$ is used to describe the multiplexing of resource blocks at an RSU when the network slicing technology is adopted for 5G autonomous vehicular networks. Notably, ${{a}_{kg}}\in \left[ 0,1 \right),\text{ }1\le k\le {{N}_{RSU}}$ and $\text{ }1\le g\le {{N}_{RSU}}$, is the element of the matrix ${{\bf{{\rm A}}}_{{N_{RSU}}*{N_{RSU}}}}$. When $k\ne g$, ${{a}_{kg}}$ is the percentage of resource blocks at the $g-th$ RSU that is scheduled for the $k-th$ RSU. When $k=g$, ${{a}_{kg}}$ is the percentage of resource blocks at the $g-th$ RSU used for local service slices. To ensure that local service slice handling occurs at the local RSU, an upper bound ratio ${{a}_{kg}^{Max}}$ is established to restrict the percentage of local RSU resource blocks used for other RUSs, i.e., $0\le {{a}_{kg}}\le {{a}_{kg}^{Max}}<1$ when $k=g$. In the proposed 5G autonomous vehicular network architecture, the RSU virtualization layer of $RS{{U}_{k}}$ can schedule the resource blocks from adjacent RSUs for local service slicing and reduce the handling latency of the service slice. The handling latency of a service slice at $In{{p}_{k}}$ is denoted as $Ts_{k}^{q}\left( k=1,2,...,{{N}_{RSU}} \right)$ and the handling latency of service slice at the local RSU $RS{{U}_{k}}$ is denoted by $Ts{{_{_{M,k}}^{q}}}$. When network slicing is adopted for 5G autonomous vehicular networks, the handling latency $Tt_{_{M}}^{q}$ in the queue of the RSU is calculated based on the total handling latency of a service slice at the RSU $RS{{U}_{k}}$, i.e., $Tt_{_{M}}^{q}=T{{s}_{M,k}}$. Moreover, the total handling latency of a service slice at $RS{{U}_{k}}$ is derived as follows:

\[T{{s}_{M,k}}\text{=}\frac{{{a}_{kk}}}{\sum\limits_{g=1}^{{{N}_{RSU}}}{{{a}_{kg}}}}Ts{{_{_{M,k}}^{q}}}+\frac{\sum\limits_{g\left( g\ne k \right)}{{{a}_{kg}}}}{\sum\limits_{g=1}^{{{N}_{RSU}}}{{{a}_{kg}}}}Ts_{_{k}}^{q}.\tag{21}\]

Based on the proposed autonomous vehicular network architecture in Fig. 8, the resource blocks in a 5G autonomous vehicular network can be multiplexed by the following process. When a service slice arrives at  $RS{{U}_{k}}$, the associated $In{{p}_{k}}$ determines whether the local resource blocks of $RS{{U}_{k}}$ can satisfy the latency requirement of service slice $T_{S}^{req}$. Additional, when the local resource blocks can satisfy the latency requirement of the service slice, i.e., $T<T_{S}^{req}$, $In{{p}_{k}}$ decreases the number of local resource blocks that can be scheduled by $RS{{U}_{k}}$ until $T=T_{S}^{req}$. In this case, the number of occupied resource blocks is $R{{B}_{rsu}}{{\left( k \right)}^{\prime }}$, which satisfies $T=T_{S}^{req}$. The remaining resource blocks $R{{B}_{rem}}\left( k \right)=R{{B}_{rsu}}\left( k \right)-R{{B}_{rsu}}{{\left( k \right)}^{\prime }}$ at $RS{{U}_{k}}$ are added to the virtual network resource pool which can be scheduled by $In{{p}_{k}}$. When the local resource blocks can not satisfy the latency requirement of a service slice, i.e., $T>T_{S}^{req}$, $In{{p}_{k}}$ increases the additional number of resource blocks scheduled by $RS{{U}_{k}}$ until $T=T_{S}^{req}$. In this case, $In{{p}_{k}}$ schedules the number of resource blocks $R{{B}_{req}}\left( k \right)=R{{B}_{rsu}}{{\left( k \right)}^{\prime }}-R{{B}_{rsu}}\left( k\right)$ from the virtual network resource pool to support the service slice handling at $RS{{U}_{k}}$. The algorithm is described in detail in the Network Slicing Algorithm.

\begin{algorithm}[t]
\caption{Network Slicing Algorithm} 
\hspace*{0.02in} {\bf Begin:} 
\begin{algorithmic}[1]
\State Initialize ${{\bf{{\rm A}}}_{{N_{RSU}}*{N_{RSU}}}}$, ${{N}_{RSU}}$, $R{{B}_{rsu}}\left( k \right)$, $R{{B}_{rem}}$, $R{{B}_{req}}$, $T_{S}^{req}$, ${{a}_{kg}^{Max}}$; 
\For{ k = 1:1: ${{N}_{RSU}}$} 
　　\State $T=Tt_{_{M}}^{p}+Tt_{_{M}}^{q}$;
　　\While{$T<T_{S}^{req}$} 
　　    \State  update $R{{B}_{rsu}}{{\left( k \right)}^{\prime }}$ by decreasing $R{{B}_{rsu}}\left( k \right)$;
        \State  calculate $T$ with $R{{B}_{rsu}}{{\left( k \right)}^{\prime }}$ by (11);
    \EndWhile
    \State  $R{{B}_{rem}}\left( k \right)=R{{B}_{rsu}}\left( k \right)-R{{B}_{rsu}}{{\left( k \right)}^{\prime }}$;
    \State  $R{{B}_{rem}}=R{{B}_{rem}}+R{{B}_{rem}}\left( k \right)$;
    \While{$T>T_{S}^{req}$} 
　　    \State  update $R{{B}_{rsu}}{{\left( k \right)}^{\prime }}$ by increasing $R{{B}_{rsu}}\left( k\right)$;
        \State  calculate $T{{s}_{M,k}}$ with $R{{B}_{rsu}}{{\left( k \right)}^{\prime }}$ by (21);
    \EndWhile
    \State  $R{{B}_{req}}\left( k \right)=R{{B}_{rsu}}{{\left( k \right)}^{\prime }}-R{{B}_{rsu}}\left( k \right)$;
    \State  $R{{B}_{req}}=R{{B}_{req}}+R{{B}_{req}}\left( k \right)$;
\EndFor
\While{$R{{B}_{rem}}\And R{{B}_{req}}\ne 0$} 
　　\State  schedule resource blocks from the virtual network resource pool to the RSU, which requires the resource blocks to reduce the handling latency of the service slice;
    \State  update the matrix ${{\bf{{\rm A}}}_{{N_{RSU}}*{N_{RSU}}}}$;
\EndWhile
\For{ k = 1:1: ${{N}_{RSU}}$}
    \State  calculate $T{{s}_{M,k}}$ by (21);
\EndFor
\end{algorithmic}
\end{algorithm}

\section{Simulation Results and Discussion}
To analyze the performance of the network slicing algorithm for 5G autonomous vehicular networks, the default simulation parameters in Section III are used in the following simulations.
When service slicing is adopted in 5G autonomous vehicular networks, all messages are classified into three types: SRSS, EDSS and EASS. Different types of service slices have different data packet sizes. Different data packet sizes have different transmission times that correspond to different slot values ${{t}_{slot}}$. To simplify the simulation analysis, the values of ${{t}_{slot}}$ are set to 25, 50 and 100 milliseconds, which correspond to the SRSS, EDSS and EASS in 5G autonomous vehicular networks with network slicing, respectively. Because SRSSs are regularly transmitted by vehicles, the number of SRSSs is obviously larger than the numbers of EDSSs and EASSs. Without loss of generality, the proportion of the numbers of SRSS, EDSS and EASS is configured as 8:1:1.
The network slicing solution is adopted for 5G autonomous vehicular networks, and the reliability and latency are then analyzed and compared based on MC simulations in this section.

Fig. 9 compares the reliability and latency utility functions in 5G autonomous vehicular networks with and without network slicing. When network slicing is adopted in 5G autonomous vehicular networks, the reliability utility function increases and the latency utility function decreases as the vehicle density increases. A conflict still occurs between the reliability and latency utility functions. Compared with the values of reliability and latency utility functions without network slicing, the reliability and latency utility function values are both improved by adopting the network slicing algorithm in 5G autonomous vehicular networks.

\begin{figure}
\vspace{0.1in}
\centerline{\includegraphics[width=9cm,draft=false]{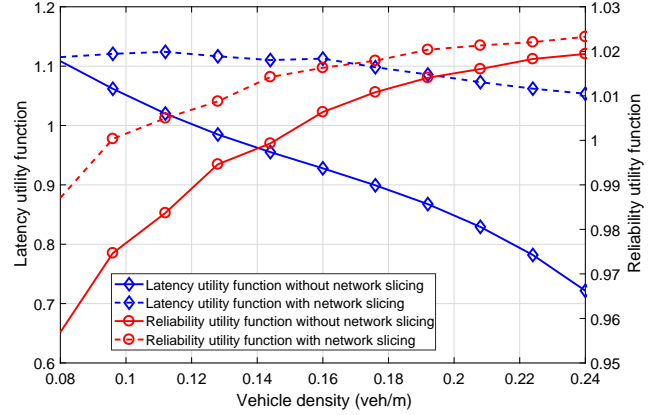}}
\caption{\small Reliability and latency utility functions in 5G autonomous vehicular networks with and without network slicing.}
\end{figure}

Fig. 10 analyzes the reliability and latency joint function with respect to the density of vehicle considering different upper bound ratios ${{a}_{kg}^{Max}}$ in 5G autonomous vehicular networks with and without network slicing. When the upper bound ratio is fixed, the values of the reliability and latency joint function with network slicing are larger than those without network slicing. When the network slicing solution is adopted and the vehicle density is less than 0.12 veh/m, the reliability and latency joint function values for ${{a}_{kg}^{Max}} = 0.8$ are larger than those for ${{a}_{kg}^{Max}} = 0.9$. If a high value of upper bound ratio, e.g., ${{a}_{kg}^{Max}} = 0.9$ is configured for low vehicle density scenarios, i.e., the vehicle density is less than 0.12 veh/m, overmuch resource blocks will be scheduled by adjacent RSUs and then the reliability and latency joint function of local RSU has to be obviously reduced. As a consequence, the reliability and latency joint function of autonomous vehicular networks is decreased.
When the network slicing solution is adopted, and the vehicle density is larger than or equal to 0.12 veh/m, the reliability and latency joint function values for ${{a}_{kg}^{Max}} = 0.8$ are less than or equal to those for ${{a}_{kg}^{Max}} = 0.9$.  If a low value of upper bound ratio, e.g., ${{a}_{kg}^{Max}} = 0.8$ is configured for the high vehicle density scenarios, i.e., the vehicle density is larger than or equal to 0.12 veh/m, a few of resource blocks can be scheduled by adjacent RSUs and then the reliability and latency joint function of adjacent RSUs has to be suppressed. As a result, the reliability and latency joint function of autonomous vehicular networks is repressed.

\begin{figure}
\vspace{0.1in}
\centerline{\includegraphics[width=9.2cm,draft=false]{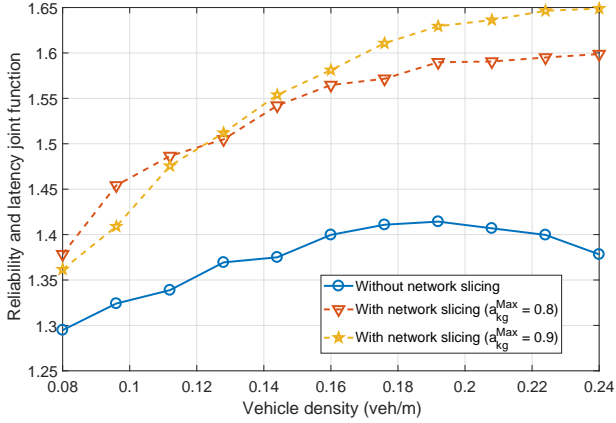}}
\caption{\small Reliability and latency joint function with respect to the density of vehicle considering different upper bound ratios ${{a}_{kg}^{Max}}$ in 5G autonomous vehicular networks with and without network slicing.}
\end{figure}

\section{Conclusion}

In this paper a reliability and latency joint function is proposed to evaluate the joint impact of reliability and latency on 5G autonomous vehicular networks. Moreover, the interactions between reliability and latency are quantitatively analyzed based on MC simulations. The simulation results indicate that a maximum reliability and latency joint function value occurs with respect to the vehicle density in 5G autonomous vehicular networks. To improve both the reliability and latency in 5G autonomous vehicular networks, i.e., to implement URLLC, a network slicing solution is proposed in this paper. Moreover, we propose a new vehicular network architecture in which network slicing is extended from network resource slicing to service and function slicing. Furthermore, a new network slicing algorithm is developed to implement URLLC in 5G autonomous vehicular networks. The simulation results show that the proposed network slicing algorithm can improve both the reliability and latency of 5G autonomous vehicular networks. Therefore, our results indicate that network slicing technology can support URLLC in 5G autonomous vehicular networks. In future work, ultra-reliable low-latency performance must be optimized in future autonomous vehicular networks, and the follow potential topics should be addressed: 1) optimize the service slicing scheme to match the type and number of messages in future autonomous vehicular networks; 2) optimize the network resource slicing scheme to account for the service slicing results; and 3) optimize the function slicing scheme to support the service and network resource slicing schemes in future autonomous vehicular networks.

\section*{Appendix A}
The message interval time generated from a vehicle is denoted as ${{T}_{0}}$. For the time duration $t\ge 0,s\ge 0$, we have the following result can be obtained:

\[P\left\{ {{T_0} > t + s|{T_0} > s} \right\} = P\left\{ {{T_0} > t} \right\} = {e^{ - {\lambda _s}t}}.\tag{22}\]

Assuming that there are $n$ vehicles in the coverage of a RSU, when the message arrival time interval at the RSU is denoted by $T_{RSU}^{s}$, the probability that $T_{RSU}^{s}$ is less than or equal to the time duration $t$ is derived as follows:

\[\begin{array}{*{20}{l}}
  {P\left\{ {T_{RSU}^s \leqslant t} \right\} = 1 - P\left\{ {T_{RSU}^s > t} \right\}} \\
  \begin{gathered}
  {\kern 1pt} {\kern 1pt}  = 1 - P\left\{ {{\text{There}}\;{\text{are}}\;{\text{no}}\;{\text{messages}}\;{\text{generated}}\;\;} \right. \hfill \\
  {\kern 1pt} {\kern 1pt} {\kern 1pt} {\kern 1pt} {\kern 1pt} {\kern 1pt} {\kern 1pt} {\kern 1pt} {\kern 1pt} {\kern 1pt} {\kern 1pt} {\kern 1pt} {\kern 1pt} {\kern 1pt} {\kern 1pt} {\kern 1pt} {\kern 1pt} {\kern 1pt} {\kern 1pt} {\kern 1pt} {\kern 1pt} {\kern 1pt} {\kern 1pt} {\kern 1pt} {\kern 1pt} {\kern 1pt} {\kern 1pt} {\kern 1pt} {\kern 1pt} {\kern 1pt} {\kern 1pt} {\kern 1pt} {\kern 1pt} {\kern 1pt} {\kern 1pt} {\kern 1pt} {\kern 1pt} {\kern 1pt} {\kern 1pt} {\kern 1pt} {\kern 1pt} {\kern 1pt} {\kern 1pt} {\text{in}}\;{\text{the}}{\kern 1pt} {\kern 1pt} {\kern 1pt} {\text{covergae}}\;{\text{area}}\;{\text{of}}\;{\text{the}} \hfill \\
\end{gathered}  \\
  {\quad \quad \quad \left. {\;{\kern 1pt} {\kern 1pt} {\kern 1pt} {\kern 1pt} {\text{RSU}}\;{\text{within}}\;{\text{the}}\;{\text{time}}\;{\text{slot}}\;\left( {0,t} \right]} \right\}} \\
  { = 1 - \frac{{\sum\limits_{n = 1}^\infty  {{{\left( {{e^{ - {\lambda _s}t}}} \right)}^n}\frac{{{{\left( {{\rho _i}L} \right)}^n}}}{{n!}}{e^{ - {\rho _i}L}}} }}{{1 - {e^{ - {\rho _i}L}}}}}
\end{array}.\tag{23}\]
Hence, the CDF of the message arrival time interval at the RSU is derived as follows:

\[F\left( t \right) = 1 - \frac{{{e^{ - {\rho _i}L\left( {1 - {e^{ - {\lambda _s}t}}} \right)}} - {e^{ - {\rho _i}L}}}}{{1 - {e^{ - {\rho _i}L}}}}.\tag{24}\]
Additional the PDF of the message arrival time interval at the RSU is derived as follows:

\[f\left( t \right) = \frac{{{\lambda _s}{\rho _i}L{e^{ - {\rho _i}L + {\rho _i}L{e^{ - {\lambda _s}t}} - {{\lambda _s}t}}}}}{{1 - {e^{ - {\rho _i}L}}}}.\tag{25}\]
Thus, Theorem 1 is proven.

\section*{Appendix B}
Assume that the message handling process at the RSU is a $GI/M/1/\infty $ queuing system. Thus, $p_{j}^{-}$ is the probability that there are $j$ messages in the queue of the RSU. Based on the results in \cite{30Medhi03}, $\left\{ p_{j}^{-}=0,j\ge 0 \right\}$ is not a stable distribution in the $GI/M/1/\infty $ queuing system when $\frac{\lambda }{c\mu }\ge 1$. $\left\{ p_{j}^{-}\ne 0,j\ge 0 \right\}$ is a stable distribution in the $GI/M/1/\infty $ queuing system when $\frac{\lambda }{c\mu }<1$. Considering $\frac{\lambda }{c\mu }<1$, the probability $p_{j}^{-}$ is expressed as follows:

\[\begin{array}{*{20}{c}}
{p_j^ -  = \sum\limits_{i = 0}^\infty  {p_i^ -  \cdot {p_{ij}}\left( 1 \right),} }&{j = 0,1,2, \cdots }
\end{array},\tag{26}\]
where ${{p}_{ij}}\left( 1 \right)$ is the one step transition probability when the number of queuing messages changes from $i$ to $j$. Considering $\sum\limits_{j=0}^{\infty }{p_{j}^{-}}=1$ and $j\ge c$, we obtain the following result:

\[{p_{ij}}\left( 1 \right) = \left\{ {\begin{array}{*{20}{c}}
{\int_0^\infty  {{e^{ - c\mu t}}\frac{{{{\left( {c\mu t} \right)}^{i - j + 1}}}}{{\left( {i - j + 1} \right)!}}dF\left( t \right),} }&{i \ge j - 1}\\
{0,}&{i < j - 1}
\end{array}} \right..\tag{27}\]
When (27) is substituted into (26) and $j\ge c$, the probability $p_{j}^{-}$ is derived as follows:

\[\begin{gathered}
  p_j^ -  = \sum\limits_{i = j - 1}^\infty  {p_i^ -  \cdot \int_0^\infty  {{e^{ - c\mu t}}\frac{{{{\left( {c\mu t} \right)}^{i - j + 1}}}}{{\left( {i - j + 1} \right)!}}dF\left( t \right)} ,}  \hfill \\
  {\kern 1pt} {\kern 1pt} {\kern 1pt} {\kern 1pt} {\kern 1pt} {\kern 1pt} {\kern 1pt} {\kern 1pt} {\kern 1pt} {\kern 1pt} {\kern 1pt} {\kern 1pt} {\kern 1pt} {\kern 1pt} {\kern 1pt} {\kern 1pt} {\kern 1pt} {\kern 1pt} {\kern 1pt} {\kern 1pt} {\kern 1pt} {\kern 1pt} {\kern 1pt} {\kern 1pt} {\kern 1pt} {\kern 1pt} {\kern 1pt} {\kern 1pt} {\kern 1pt} {\kern 1pt} {\kern 1pt} {\kern 1pt} {\kern 1pt} {\kern 1pt} {\kern 1pt} {\kern 1pt} {\kern 1pt} {\kern 1pt} {\kern 1pt} {\kern 1pt} {\kern 1pt} {\kern 1pt} {\kern 1pt} {\kern 1pt} {\kern 1pt} {\kern 1pt} {\kern 1pt} {\kern 1pt} {\kern 1pt} {\kern 1pt} {\kern 1pt} {\kern 1pt} {\kern 1pt} {\kern 1pt} {\kern 1pt} {\kern 1pt} {\kern 1pt} {\kern 1pt} {\kern 1pt} {\kern 1pt} {\kern 1pt} {\kern 1pt} {\kern 1pt} {\kern 1pt} {\kern 1pt} {\kern 1pt} {\kern 1pt} {\kern 1pt} {\kern 1pt} {\kern 1pt} {\kern 1pt} {\kern 1pt} {\kern 1pt} {\kern 1pt} {\kern 1pt} {\kern 1pt} {\kern 1pt} {\kern 1pt} {\kern 1pt} {\kern 1pt} {\kern 1pt} {\kern 1pt} {\kern 1pt} {\kern 1pt} {\kern 1pt} {\kern 1pt} {\kern 1pt} {\kern 1pt} {\kern 1pt} {\kern 1pt} j = c,c + 1,c + 2, \cdots  \hfill \\
\end{gathered} .\tag{28}\]

Based on the results in \cite{30Medhi03}, the following relation can be derived:

\[\sum\limits_{k = 0}^\infty  k  \cdot \int_0^\infty  {{e^{ - c\mu t}}} \frac{{{{\left( {c\mu t} \right)}^k}}}{{k!}}dF\left( t \right) = \frac{{c\mu }}{\lambda } > 0.\tag{29}\]
Additionally, based on (29), the following equation can be derived:

\[\sum\limits_{k = 0}^\infty  {{\delta ^k}}  \cdot \int_0^\infty  {{e^{ - c\mu t}}} \frac{{{{\left( {c\mu t} \right)}^k}}}{{k!}}dF\left( t \right) = f\left( {c\mu \left( {1 - \delta } \right)} \right) = \delta .\tag{30}\]

Based on the results in \cite{30Medhi03}, (30) can be solved, and the solution is exclusively in the range of $\left( 0,1 \right)$. When the solution of (30) is denoted as $\delta \left( 0<\delta <1 \right)$, the probability $p_{j}^{-}$ is calculated as follows:

\[\begin{array}{*{20}{c}}
{p_j^ -  = {K^*} \cdot {\delta ^{j - c}},}&{j = c - 1,c,c + 1, \cdots }
\end{array},\tag{31}\]
where ${{K}^{*}}$ is a constant. When $j=c-1$, ${{K}^{*}}$ is calculated as follows:

\[{K^*} = \delta  \cdot p_{c - 1}^ - .\tag{32}\]

Considering $c=1$, i.e., the $GI/M/1/\infty $ queuing system is implemented, ${{K}^{*}}$ can be simply denoted by ${{K}^{*}}=\delta \cdot p_{0}^{-}$. Furthermore, the probability $p_{j}^{-}$ is derived as follows:

\[\begin{array}{*{20}{c}}
{p_j^ -  = p_0^ -  \cdot {\delta ^j},}&{j = 0,1,2, \cdots }
\end{array} .\tag{33}\]

Considering $\sum\limits_{j=0}^{\infty }{p_{j}^{-}}=1$, the probability of an empty queue is calculated by $p_{0}^{-}=1-\delta $. By substituting the value of $p_{0}^{-}$ into (33), the probability $p_{j}^{-}$ can be calculated as follows:
\[\begin{array}{*{20}{c}}
{p_j^ -  = \left( {1 - \delta } \right) \cdot {\delta ^j},}&{j = 0,1,2, \cdots }
\end{array}.\tag{34}\]
When $c$ is configured as $c=1$, ${{K}^{*}}$ is calculated as follows:

\[{K^*} = \delta  \cdot \left( {1 - \delta } \right).\tag{35}\]

In this paper the message is assumed to be serviced by the first-in-first-out (FIFO) scheme in the $GI/M/c/\infty $ queuing system. The waiting time of the $m-th$ message is denoted by ${{W}_{{{q}_{m}}}}$ in the $GI/M/c/\infty $ queuing system. When $\frac{\lambda }{c\mu }<1$, the distribution of ${{W}_{{{q}_{m}}}}$ is expressed as $P\left\{ {{W}_{{{q}_{m}}}}\le t \right\}={{W}_{{{q}_{m}}}}\left( t \right)$. When the queuing length is $N_{m}^{-}=j$ and $j<c$, the $m-th$ message can be directly handled without waiting. When $j\ge c$, the $m-th$ message must be wait until the services for $j-c+1$ messages have been completed in the queuing system. Based on the total probability theorem, the waiting time of the $m-th$ message is expressed as follows:

\[\begin{gathered}
  {W_{{q_m}}}\left( t \right) = \sum\limits_{j = 0}^{c - 1} {P\left\{ {N_m^ -  = j} \right\} \cdot 0}  \hfill \\
  {\kern 1pt} {\kern 1pt} {\kern 1pt} {\kern 1pt} {\kern 1pt} {\kern 1pt} {\kern 1pt} {\kern 1pt} {\kern 1pt} {\kern 1pt} {\kern 1pt} {\kern 1pt} {\kern 1pt} {\kern 1pt} {\kern 1pt} {\kern 1pt} {\kern 1pt} {\kern 1pt} {\kern 1pt} {\kern 1pt} {\kern 1pt} {\kern 1pt} {\kern 1pt} {\kern 1pt} {\kern 1pt} {\kern 1pt} {\kern 1pt} {\kern 1pt} {\kern 1pt} {\kern 1pt} {\kern 1pt} {\kern 1pt} {\kern 1pt} {\kern 1pt} {\kern 1pt} {\kern 1pt} {\kern 1pt} {\kern 1pt} {\kern 1pt} {\kern 1pt} {\kern 1pt} {\kern 1pt}  + \sum\limits_{j = c}^\infty  {P\left\{ {N_m^ -  = j} \right\}}  \cdot \int_0^t {{e^{ - c\mu x}}} \frac{{{{\left( {c\mu x} \right)}^{j - c}} \cdot c\mu }}{{\left( {j - c} \right)!}}dx \hfill \\
\end{gathered} .\tag{36}\]

Considering $\frac{\lambda }{c\mu }<1$, the limitation of the waiting time of the $m-th$ message is derived as follows:

\[\begin{array}{*{20}{l}}
  \begin{gathered}
  \mathop {\lim }\limits_{m \to \infty } {W_{{q_m}}}\left( t \right) = \sum\limits_{j = 0}^{c - 1} {p_j^ -  \cdot 0}  \hfill \\
  {\kern 1pt} {\kern 1pt} {\kern 1pt} {\kern 1pt} {\kern 1pt} {\kern 1pt} {\kern 1pt} {\kern 1pt} {\kern 1pt} {\kern 1pt} {\kern 1pt} {\kern 1pt} {\kern 1pt} {\kern 1pt} {\kern 1pt} {\kern 1pt} {\kern 1pt} {\kern 1pt} {\kern 1pt} {\kern 1pt} {\kern 1pt} {\kern 1pt} {\kern 1pt} {\kern 1pt} {\kern 1pt} {\kern 1pt} {\kern 1pt} {\kern 1pt} {\kern 1pt} {\kern 1pt} {\kern 1pt} {\kern 1pt} {\kern 1pt} {\kern 1pt} {\kern 1pt} {\kern 1pt} {\kern 1pt} {\kern 1pt} {\kern 1pt} {\kern 1pt} {\kern 1pt} {\kern 1pt} {\kern 1pt} {\kern 1pt} {\kern 1pt} {\kern 1pt} {\kern 1pt} {\kern 1pt} {\kern 1pt} {\kern 1pt} {\kern 1pt} {\kern 1pt} {\kern 1pt} {\kern 1pt} {\kern 1pt} {\kern 1pt} {\kern 1pt} {\kern 1pt} {\kern 1pt}  + \sum\limits_{j = c}^\infty  {p_j^ -  \cdot \int_0^t {{e^{ - c\mu x}}} \frac{{c\mu {{\left( {c\mu x} \right)}^{j - c}}}}{{\left( {j - c} \right)!}}dx}  \hfill \\
\end{gathered}  \\
  {\qquad \qquad \qquad  = {K^*}\sum\limits_{j = c}^\infty  {{\delta ^{j - c}}}  \cdot \int_0^t {{e^{ - c\mu x}}} \frac{{c\mu {{\left( {c\mu x} \right)}^{j - c}}}}{{\left( {j - c} \right)!}}dx} \\
  {\qquad \qquad \qquad  = \frac{{{K^*}}}{{1 - \delta }}\left[ {1 - {e^{ - c\mu \left( {1 - \delta } \right)t}}} \right]}
\end{array}.\tag{37}\]

The service time of the $m-th$ message is denoted by ${{\chi }_{m}}$ and the distribution of ${{\chi }_{m}}$ is denoted by ${{\chi }_{m}}\left( t \right)$. Hence, the distribution of the message dwelling time is expressed as $W\left( t \right)=\underset{m\to \infty }{\mathop{\lim }}\,\left( {{\chi }_{m}}\left( t \right)+{{W}_{{{q}_{m}}}}\left( t \right) \right)$. Because ${{W}_{{{q}_{m}}}}$ and ${{\chi }_{m}}$ are independent, the distribution of the message dwelling time is derived as follows:

\[W\left( t \right) = 1 - {e^{ - \mu t}} + \frac{{{K^*}}}{{1 - \delta }}\left[ {1 - {e^{ - c\mu \left( {1 - \delta } \right)t}}} \right].\tag{38}\]
Thus, Theorem 2 is proven.

\section*{Appendix C}

Given that the density of InP is ${{\lambda }_{Inp}}$, a typical PVT cell area denoted by ${{\mathbb{R}}_{Inp}}$, follows a Gamma distribution \cite{36Xiang13}. The PDF of the InP distribution is expressed as follows:
\[{{f}_{{{\mathbb{R}}_{Inp}}}}\left( x \right)=\frac{{{\left( b{{\lambda }_{Inp}} \right)}^{a}}}{\Gamma \left( a \right)}{{x}^{a-1}}{{e}^{-bx{{\lambda }_{Inp}}}},\tag{39}\]
where $\Gamma \left( x \right)=\int_{0}^{\infty }{{{t}^{x-1}}}{{e}^{-t}}dt$ is a Gamma function, $a$ is the shape parameter and $b{{\lambda }_{Inp}}$ is the inverse scale parameter of the Gamma distribution. The default values of $a$ and $b$ are configured as $a=3.61$ and $b=3.57$ in the simulation analysis, respectively. The road length in a coverage area of an InP is derived as follows:

\[{{L}_{{{\mathbb{R}}_{Inp}}}}\left( x \right)=\int_{0}^{\infty }{{{\rho }_{road}}\frac{{{\left( b{{\lambda }_{Inp}} \right)}^{a}}}{\Gamma \left( a \right)}{{x}^{a-1}}{{e}^{-bx{{\lambda }_{Inp}}}}dx},\tag{40}\]
where ${{\rho }_{road}}$ is the road density in an urban environment. Considering the uniform distribution of RSUs along the road, the expected ${{N}_{RSU}}$ is derived as follows:

\[{\rm E}\left( {{N_{RSU}}} \right) = \frac{{\int_0^\infty  {{\rho _{road}}\frac{{{{\left( {b{\lambda _{Inp}}} \right)}^a}}}{{\Gamma \left( a \right)}}{x^{a - 1}}{e^{ - bx{\lambda _{Inp}}}}dx} }}{{{\rho _{RSU}}}}.\tag{41}\]




\begin{IEEEbiography}[{\includegraphics[width=0.8in,height=1in,clip,keepaspectratio]{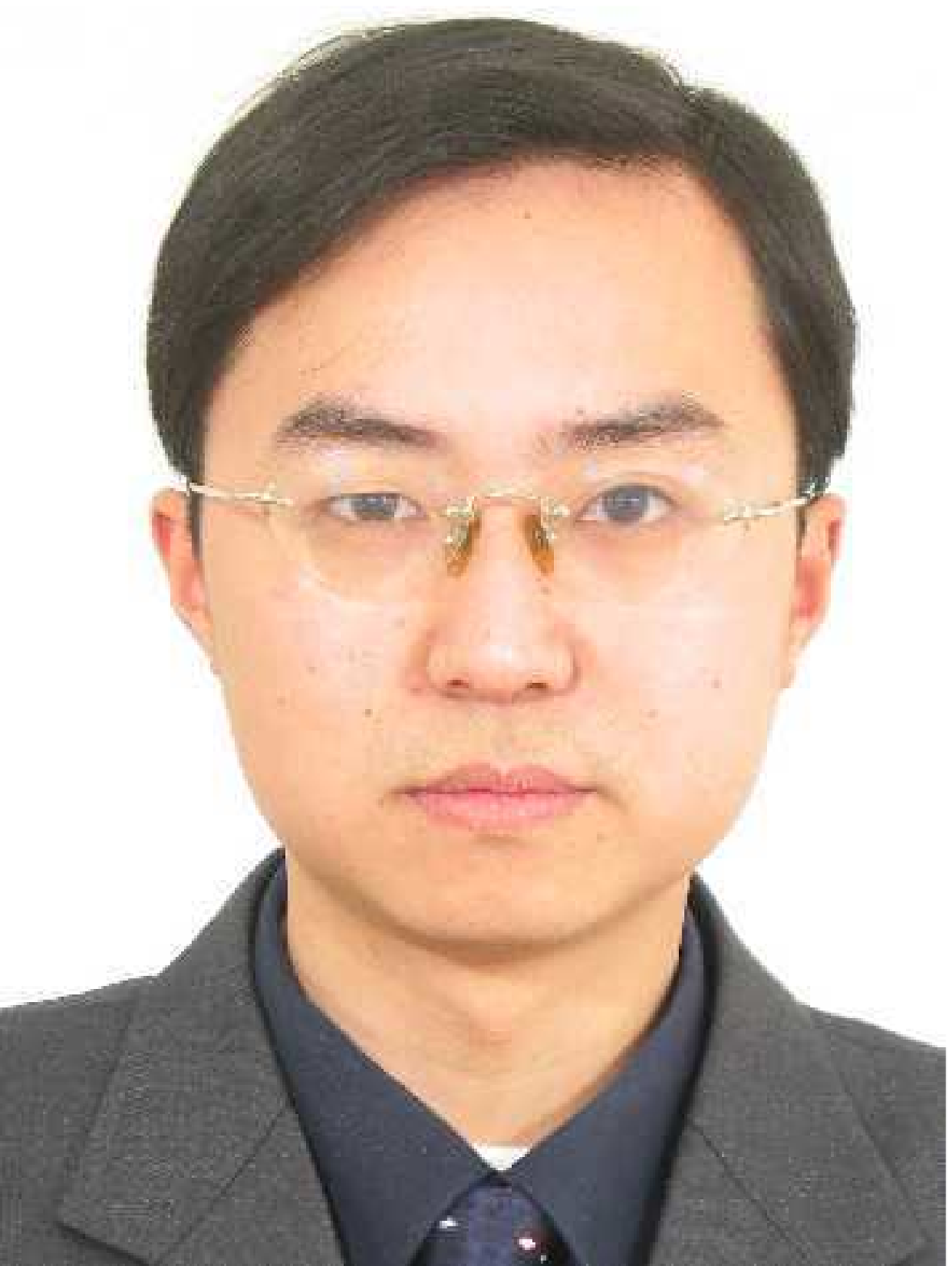}}]{Xiaohu~Ge}
(M'09-SM'11) received the Ph.D. degree in communication and information engineering from the Huazhong University of Science and Technology (HUST),Wuhan, China, in 2003. Since 2005, he has been with HUST, where he is currently a Full Professor with the School of Electronic Information and Communications. He is an Adjunct Professor with Faculty of Engineering and Information Technology, University of Technology Sydney, Ultimo, NSW, Australia. He is the Director of the China International Joint Research Center of Green Communications and Networking.
\end{IEEEbiography}
\vspace{-6 mm}

\end{document}